\documentclass[12pt]{article}
\usepackage{amssymb}

\usepackage{epsfig}

\begin{document}

\title{Poisson-Vlasov : Stochastic representation and numerical codes}
\author{Elena Floriani\thanks{%
Centre de Physique Th\'{e}orique, CNRS Luminy, case 907, F-13288 Marseille
Cedex 9, France; floriani@cpt.univ-mrs.fr, lima@cpt.univ-mrs.fr}, Ricardo
Lima\footnotemark \ and R. Vilela Mendes\thanks{%
Centro de Fus\~{a}o Nuclear - EURATOM/IST Association, Instituto Superior
T\'{e}cnico, Av. Rovisco Pais 1, 1049-001 Lisboa, Portugal} \thanks{%
CMAF, Complexo Interdisciplinar, Universidade de Lisboa, Av. Gama Pinto, 2 -
1649-003 Lisboa (Portugal), e-mail: vilela@cii.fc.ul.pt;
http://label2.ist.utl.pt/vilela/}}
\date{}
\maketitle

\begin{abstract}
A stochastic representation for the solutions of the Poisson-Vlasov
equation, with several charged species, is obtained. The representation
involves both an exponential and a branching process and it provides an
intuitive characterization of the nature of the solutions and its
fluctuations. Here, the stochastic representation is also proposed as a tool
for the numerical evaluation of the solutions
\end{abstract}

\section{\textbf{Introduction}}

It is well known that the solutions of linear elliptic and parabolic
equations, both with Cauchy and Dirichlet boundary conditions, have a
probabilistic interpretation. This is a very classical field which may be
traced back to the work of Courant, Friedrichs and Lewy \cite{Courant} in
the 20's. In spite of the pioneering work of McKean \cite{McKean}, the
question of whether useful probabilistic representations could also be found
for a large class of nonlinear equations remained an essentially open
problem for many years. It was only in the 90's that, with the work of Dynkin%
\cite{Dynkin1} \cite{Dynkin2}, such a theory started to take shape. For
nonlinear diffusion processes, the branching exit Markov systems, that is,
processes that involve both diffusion and branching, seem to play the same
role as Brownian motion in the linear equations. However the theory is still
limited to some classes of nonlinearities and there is much room for further
mathematical improvement.

Another field, where considerable recent advances were achieved, was the
probabilistic representation of the Fourier transformed Navier-Stokes
equation, first with the work of LeJan and Sznitman\cite{Jan}, later
followed by extensive developments of the Oregon school\cite{Waymire} \cite
{Bhatta1} \cite{Ossiander}. In all cases the stochastic representation
defines a process fort which the mean values of some functionals coincide
with the solution of the deterministic equation.

Stochastic representations, in addition to its intrinsic mathematical
relevance, have several practical implications:

(i) They provide an intuitive characterization of the equation solutions;

(ii) By the study of exit times from a domain they sometimes provide access
to quantities that cannot be obtained by perturbative methods\cite{QCD}

(iii) They provide a calculation tool which may replace, for example, the
need for very fine integration grids at high Reynolds numbers;

(iv) By associating a stochastic process to the solutions of the equation,
they may also provide an intrinsic characterization of the nature of the
fluctuations associated to the physical system. In some cases the stochastic
process is essentially unique, in others there is a class of processes with
means leading to the same solution.

In \cite{Vilela1} a stochastic representation has been obtained for the
solutions of the Fourier-transformed Poisson-Vlasov equation in 3 dimensions
for particles of one charge species on an arbitrary background. Here this
result is generalized for the case of several charged species. As before the
representation involves both an exponential and a branching process, the
solution being obtained from the expectation value of a multiplicative
functional over backwards in time realizations of the process.

The backwards in time realization of the process turns out to be appropriate
for (parallelizable) numerical evaluation of the solutions and the Fourier
representation adequate to obtain information on the small scale behaviour.

\section{The stochastic representations}

Consider a multi-species Poisson-Vlasov equation in 3+1 space-time
dimensions 
\begin{equation}
\frac{\partial f_{i}}{\partial t}+\stackrel{\rightarrow }{v}\cdot \nabla
_{x}f_{i}-\frac{e_{i}}{m_{i}}\nabla _{x}\Phi \cdot \nabla _{v}f_{i}=0
\label{2.1}
\end{equation}
$\left( i=1,2\right) $, with 
\begin{equation}
\Delta _{x}\Phi =-4\pi \left\{ \sum_{i}e_{i}\int f_{i}\left( \stackrel{%
\rightarrow }{x},\stackrel{\rightarrow }{v},t\right) d^{3}v\right\}
\label{2.2}
\end{equation}

Passing to the Fourier transform 
\begin{equation}
F_{i}\left( \xi ,t\right) =\frac{1}{\left( 2\pi \right) ^{3}}\int d^{6}\eta
f_{i}\left( \eta ,t\right) e^{i\xi \cdot \eta }  \label{2.3}
\end{equation}
with $\eta =\left( \stackrel{\rightarrow }{x},\stackrel{\rightarrow }{v}%
\right) $ and $\xi =\left( \stackrel{\rightarrow }{\xi _{1}},\stackrel{%
\rightarrow }{\xi _{2}}\right) \circeq \left( \xi _{1},\xi _{2}\right) $,
one obtains 
\begin{eqnarray}
0 &=&\frac{\partial F_{i}\left( \xi ,t\right) }{\partial t}-\stackrel{%
\rightarrow }{\xi _{1}}\cdot \nabla _{\xi _{2}}F_{i}\left( \xi ,t\right)
\label{2.4} \\
&&+\frac{4\pi e_{i}}{m_{i}}\int d^{3}\xi _{1}^{^{\prime }}F_{i}\left( \xi
_{1}-\xi _{1}^{^{\prime }},\xi _{2},t\right) \frac{\stackrel{\rightarrow }{%
\xi _{2}}\cdot \stackrel{\rightarrow }{\xi _{1}^{^{\prime }}}}{\left| \xi
_{1}^{^{\prime }}\right| ^{2}}\sum_{j}e_{j}F_{j}\left( \xi _{1}^{^{\prime
}},0,t\right)  \nonumber
\end{eqnarray}
Changing variables to 
\begin{equation}
\tau =\gamma \left( \left| \xi _{2}\right| \right) t  \label{2.5}
\end{equation}
where $\gamma \left( \left| \xi _{2}\right| \right) $ is a positive
continuous function satisfying 
\[
\begin{array}{lllll}
\gamma \left( \left| \xi _{2}\right| \right) =1 &  & \mathnormal{if} &  & 
\left| \xi _{2}\right| <1 \\ 
\gamma \left( \left| \xi _{2}\right| \right) \geq \left| \xi _{2}\right| & 
& \mathnormal{if} &  & \left| \xi _{2}\right| \geq 1
\end{array}
\]
leads to 
\begin{eqnarray}
\frac{\partial F_{i}\left( \xi ,\tau \right) }{\partial \tau } &=&\frac{%
\stackrel{\rightarrow }{\xi _{1}}}{\gamma \left( \left| \xi _{2}\right|
\right) }\cdot \nabla _{\xi _{2}}F_{i}\left( \xi ,\tau \right) -\frac{4\pi
e_{i}}{m_{i}}\int d^{3}\xi _{1}^{^{\prime }}F_{i}\left( \xi _{1}-\xi
_{1}^{^{\prime }},\xi _{2},\tau \right)  \nonumber \\
&&\times \frac{\stackrel{\rightarrow }{\xi _{2}}\cdot \stackrel{\wedge }{\xi
_{1}^{^{\prime }}}}{\gamma \left( \left| \xi _{2}\right| \right) \left| \xi
_{1}^{^{\prime }}\right| }\sum_{j}e_{j}F_{j}\left( \xi _{1}^{^{\prime
}},0,\tau \right)  \label{2.6}
\end{eqnarray}
with $\stackrel{\wedge }{\xi _{1}}=\frac{\stackrel{\rightarrow }{\xi _{1}}}{%
\left| \xi _{1}\right| }$. Eq.(\ref{2.6}) written in integral form, is 
\begin{eqnarray}
F_{i}\left( \xi ,\tau \right) &=&e^{\tau \frac{\stackrel{\rightarrow }{\xi
_{1}}}{\gamma \left( \left| \xi _{2}\right| \right) }\cdot \nabla _{\xi
_{2}}}F_{i}\left( \xi _{1},\xi _{2},0\right) -\frac{4\pi e_{i}}{m_{i}}%
\int_{0}^{\tau }dse^{\left( \tau -s\right) \frac{\stackrel{\rightarrow }{\xi
_{1}}}{\gamma \left( \left| \xi _{2}\right| \right) }\cdot \nabla _{\xi
_{2}}}  \label{2.7} \\
&&\times \int d^{3}\xi _{1}^{^{\prime }}F_{i}\left( \xi _{1}-\xi
_{1}^{^{\prime }},\xi _{2},s\right) \frac{\stackrel{\rightarrow }{\xi _{2}}%
\cdot \stackrel{\wedge }{\xi _{1}^{^{\prime }}}}{\gamma \left( \left| \xi
_{2}\right| \right) \left| \xi _{1}^{^{\prime }}\right| }\sum_{j}e_{j}F_{j}%
\left( \xi _{1}^{^{\prime }},0,s\right)  \nonumber
\end{eqnarray}

For convenience, a stochastic representation is going to be written for the
following function 
\begin{equation}
\chi _{i}\left( \xi _{1},\xi _{2},\tau \right) =e^{-\lambda \tau }\frac{%
F_{i}\left( \xi _{1},\xi _{2},\tau \right) }{h\left( \xi _{1}\right) }
\label{2.8}
\end{equation}
with $\lambda $ a constant and $h\left( \xi _{1}\right) $ a positive
function to be specified later on. The integral equation for $\chi \left(
\xi _{1},\xi _{2},\tau \right) $ is 
\begin{eqnarray}
\chi _{i}\left( \xi _{1},\xi _{2},\tau \right) &=&e^{-\lambda \tau }\chi
_{i}\left( \xi _{1},\xi _{2}+\tau \frac{\xi _{1}}{\gamma \left( \left| \xi
_{2}\right| \right) },0\right) -\frac{8\pi e_{i}}{m_{i}\lambda }\frac{\left(
\left| \xi _{1}\right| ^{-1}h*h\right) \left( \xi _{1}\right) }{h\left( \xi
_{1}\right) }\int_{0}^{\tau }ds\lambda e^{-\lambda s}  \nonumber \\
&&\times \int d^{3}\xi _{1}^{^{\prime }}p\left( \xi _{1},\xi _{1}^{^{\prime
}}\right) \chi _{i}\left( \xi _{1}-\xi _{1}^{^{\prime }},\xi _{2}+s\frac{\xi
_{1}}{\gamma \left( \left| \xi _{2}\right| \right) },\tau -s\right) 
\nonumber \\
&&\times \frac{\stackrel{\rightarrow }{\xi _{2}}\cdot \stackrel{\wedge }{\xi
_{1}^{^{\prime }}}}{\gamma \left( \left| \xi _{2}\right| \right) }\sum_{j}%
\frac{1}{2}e_{j}e^{\lambda \left( \tau -s\right) }\chi _{j}\left( \xi
_{1}^{^{\prime }},0,\tau -s\right)  \label{2.9}
\end{eqnarray}
with 
\begin{equation}
\left( \left| \xi _{1}^{^{\prime }}\right| ^{-1}h*h\right) \left( \xi
_{1}\right) =\int d^{3}\xi _{1}^{^{\prime }}\left| \xi _{1}^{^{\prime
}}\right| ^{-1}h\left( \xi _{1}-\xi _{1}^{^{\prime }}\right) h\left( \xi
_{1}^{^{\prime }}\right)  \label{2.10}
\end{equation}
and 
\begin{equation}
p\left( \xi _{1},\xi _{1}^{^{\prime }}\right) =\frac{\left| \xi
_{1}^{^{\prime }}\right| ^{-1}h\left( \xi _{1}-\xi _{1}^{^{\prime }}\right)
h\left( \xi _{1}^{^{\prime }}\right) }{\left( \left| \xi _{1}^{^{\prime
}}\right| ^{-1}h*h\right) }  \label{2.11}
\end{equation}

Eq.(\ref{2.9}) has a stochastic interpretation as an exponential process
(with a time shift in the second variable) plus a branching process. $%
p\left( \xi _{1},\xi _{1}^{^{\prime }}\right) d^{3}\xi _{1}^{^{\prime }}$ is
the probability that, given a $\xi _{1}$ mode, one obtains a $\left( \xi
_{1}-\xi _{1}^{^{\prime }},\xi _{1}^{^{\prime }}\right) $ branching with $%
\xi _{1}^{^{\prime }}$ in the volume $\left( \xi _{1}^{^{\prime }},\xi
_{1}^{^{\prime }}+d^{3}\xi _{1}^{^{\prime }}\right) $. $\chi \left( \xi
_{1},\xi _{2},\tau \right) $ is computed from the expectation value of a
multiplicative functional associated to the processes. Convergence of the
multiplicative functional hinges on the fulfilling of the following
conditions :

(A) $\left| \frac{F_{i}\left( \xi _{1},\xi _{2},0\right) }{h\left( \xi
_{1}\right) }\right| \leq 1$

(B) $\left( \left| \xi _{1}^{^{\prime }}\right| ^{-1}h*h\right) \left( \xi
_{1}\right) \leq h\left( \xi _{1}\right) $

Condition (B) is satisfied, for example, for 
\begin{equation}
h\left( \xi _{1}\right) =\frac{c}{\left( 1+\left| \xi _{1}\right|
^{2}\right) ^{2}}\hspace{1cm}\mathnormal{and}\hspace{1cm}c\leq \frac{1}{3\pi 
}  \label{2.12}
\end{equation}
Indeed computing $\left| \xi _{1}^{^{\prime }}\right| ^{-1}h*h$ one obtains 
\begin{equation}
\begin{array}{lll}
c^{2}\Gamma \left( \xi _{1}\right) =\left( \left| \xi _{1}^{^{\prime
}}\right| ^{-1}h*h\right) \left( \xi _{1}\right) & =2\pi c^{2} & \left\{ 
\frac{2\ln \left( 1+\left| \xi _{1}\right| ^{2}\right) }{\left| \xi
_{1}\right| ^{2}\left( \left| \xi _{1}\right| ^{2}+4\right) ^{2}}+\frac{1}{%
\left| \xi _{1}\right| ^{2}\left( \left| \xi _{1}\right| ^{2}+4\right) }%
\right. \\ 
&  & \left. +\frac{\left| \xi _{1}\right| ^{2}-4}{2\left| \xi _{1}\right|
^{3}\left( \left| \xi _{1}\right| ^{2}+4\right) ^{2}}\left( \frac{\pi }{2}%
-\tan ^{-1}\left( \frac{2-2\left| \xi _{1}\right| ^{2}}{4\left| \xi
_{1}\right| }\right) \right) \right\}
\end{array}
\label{2.13}
\end{equation}
Then $\frac{1}{h\left( \xi _{1}\right) }\left( \left| \xi _{1}^{^{\prime
}}\right| ^{-1}h*h\right) \left( \xi _{1}\right) $ is bounded by a constant
for all $\left| \xi _{1}\right| $, and choosing $c$ sufficiently small,
condition (B) is satisfied.

Once $h\left( \xi _{1}\right) $ consistent with (B) is found, condition (A)
only puts restrictions on the initial conditions. Now one constructs the
stochastic process $X\left( \xi _{1},\xi _{2},\tau \right) $.

Because $e^{-\lambda \tau }$ is the survival probability during time $\tau $
of an exponential process with parameter $\lambda $ and $\lambda e^{-\lambda
s}ds$ the decay probability in the interval $\left( s,s+ds\right) $, $\chi
_{i}\left( \xi _{1},\xi _{2},\tau \right) $ in Eq.(\ref{2.9}) is obtained as
the expectation value of a multiplicative functional for the following
backward-in-time process, which we denote as \textit{process I} :

Starting at $\left( \xi _{1},\xi _{2},\tau \right) $, a particle of species $%
i$ lives for an exponentially distributed time $s$ up to time $\tau -s$. At
its death a coin $l_{s}$ (probabilities $\frac{1}{2},\frac{1}{2}$) is
tossed. If $l_{s}=0$ two new particles of the same species as the original
one are born at time $\tau -s$ with Fourier modes $\left( \xi _{1}-\xi
_{1}^{^{\prime }},\xi _{2}+s\frac{\xi _{1}}{\gamma \left( \left| \xi
_{2}\right| \right) }\right) $ and $\left( \xi _{1}^{^{\prime }},0\right) $
with probability density $p\left( \xi _{1},\xi _{1}^{^{\prime }}\right) $.
If $l_{s}=1$ the two new particles are of different species. Each one of the
newborn particles continues its backward-in-time evolution, following the
same death and birth laws. When one of the particles of this tree reaches
time zero it samples the initial condition. The multiplicative functional of
the process is the product of the following contributions:

- At each branching point where two particles are born , the coupling
constant is 
\begin{equation}
g_{ij}\left( \xi _{1},\xi _{1}^{^{\prime }},s\right) =-e^{\lambda \left(
\tau -s\right) }\frac{8\pi e_{i}e_{j}}{m_{i}\lambda }\frac{\left( \left| \xi
_{1}^{^{\prime }}\right| ^{-1}h*h\right) \left( \xi _{1}\right) }{h\left(
\xi _{1}\right) }\frac{\stackrel{\rightarrow }{\xi _{2}}\cdot \stackrel{%
\symbol{94}}{\xi _{1}^{^{\prime }}}}{\gamma \left( \left| \xi _{2}\right|
\right) }  \label{2.14}
\end{equation}

- When one particle reaches time zero and samples the initial condition the
coupling is 
\begin{equation}
g_{0i}\left( \xi _{1},\xi _{2}\right) =\frac{F_{i}\left( \xi _{1},\xi
_{2},0\right) }{h\left( \xi _{1}\right) }  \label{2.16}
\end{equation}

The multiplicative functional is the product of all these couplings for each
realization of the process $X\left( \xi _{1},\xi _{2},\tau \right) $, this
process being obtained as the limit of the following iterative process 
\begin{eqnarray}
&&X_{i}^{\left( k+1\right) }\left( \xi _{1},\xi _{2},\tau \right)  \nonumber
\\
&=&\chi _{i}\left( \xi _{1},\xi _{2}+\tau \frac{\xi _{1}}{\gamma \left(
\left| \xi _{2}\right| \right) },0\right) \mathbf{1}_{\left[ s>\tau \right]
}+g_{ii}\left( \xi _{1},\xi _{1}^{^{\prime }},s\right)  \label{2.16a} \\
&&\times X_{i}^{\left( k\right) }\left( \xi _{1}-\xi _{1}^{^{\prime }},\xi
_{2}+s\frac{\xi _{1}}{\gamma \left( \left| \xi _{2}\right| \right) },\tau
-s\right) X_{i}^{\left( k\right) }\left( \xi _{1}^{^{\prime }},0,\tau
-s\right) \mathbf{1}_{\left[ s<\tau \right] }\mathbf{1}_{\left[
l_{s}=0\right] }  \nonumber \\
&&+g_{ij}\left( \xi _{1},\xi _{1}^{^{\prime }},s\right) X_{i}^{\left(
k\right) }\left( \xi _{1}-\xi _{1}^{^{\prime }},\xi _{2}+s\frac{\xi _{1}}{%
\gamma \left( \left| \xi _{2}\right| \right) },\tau -s\right) X_{j}^{\left(
k\right) }\left( \xi _{1}^{^{\prime }},0,\tau -s\right) \mathbf{1}_{\left[
s<\tau \right] }\mathbf{1}_{\left[ l_{s}=1\right] }  \nonumber
\end{eqnarray}
Then, each $\chi _{i}\left( \xi _{1},\xi _{2},\tau \right) $ is the
expectation value of the functional. 
\begin{equation}
\chi _{i}\left( \xi _{1},\xi _{2},\tau \right) =\mathbb{E}\left\{ \Pi \left(
g_{0}g_{0}^{^{\prime }}\cdots \right) \left( g_{ii}g_{ii}^{^{\prime }}\cdots
\right) \left( g_{ij}g_{ij}^{^{\prime }}\cdots \right) \right\}  \label{2.17}
\end{equation}

For example, for the realization in Fig.1 the contribution to the
multiplicative functional is 
\begin{eqnarray}
&&g_{ij}\left( \xi _{1},\xi _{1}^{^{\prime }},\tau -s_{1}\right)
g_{ji}\left( \xi _{1}-\xi _{1}^{^{\prime }},\xi _{1}^{^{\prime \prime
}},\tau -s_{2}\right) g_{ii}\left( \xi _{1}^{^{\prime }},\xi
_{1}^{^{^{\prime \prime \prime }}},\tau -s_{3}\right)  \nonumber \\
&&\times g_{0i}\left( \xi _{1}^{^{\prime }}-\xi _{1}^{^{\prime \prime \prime
}},k_{3}\right) g_{0i}\left( \xi _{1}^{^{^{\prime \prime \prime }}},0\right)
g_{0j}\left( \xi _{1}^{^{\prime \prime }},0\right) g_{0i}\left( \xi _{1}-\xi
_{1}^{^{\prime }}-\xi _{1}^{^{\prime \prime }},k_{2}\right)  \label{2.18}
\end{eqnarray}

\begin{figure}[tbh]
\begin{center}
\psfig{figure=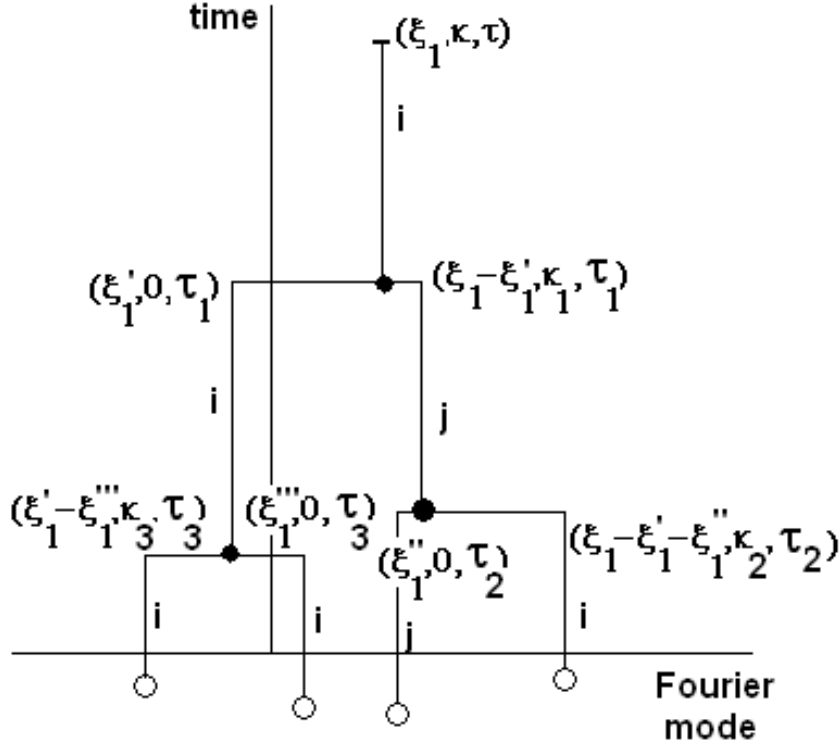,width=11truecm}
\end{center}
\caption{A sample path of the stochastic process I}
\end{figure}

and 
\begin{equation}
\begin{array}{lll}
k & = & \xi _{2} \\ 
k_{1} & = & k+\left( \tau -\tau _{1}\right) \frac{\xi _{1}}{\gamma \left(
\left| \xi _{2}\right| \right) } \\ 
k_{2} & = & k_{1}+\left( \tau _{2}-\tau _{1}\right) \frac{\left( \xi
_{1}-\xi _{1}^{^{\prime }}\right) }{\gamma \left( \left| k_{1}\right|
\right) } \\ 
k_{3} & = & \left( \tau _{3}-\tau _{1}\right) \xi _{1}^{^{\prime }}
\end{array}
\label{2.19}
\end{equation}
With the conditions (A) and (B), choosing 
\begin{equation}
\lambda \geq \left| \frac{8\pi e_{i}e_{j}}{\min_{i}\left\{ m_{i}\right\} }%
\right|  \label{2.20}
\end{equation}
and 
\begin{equation}
c\leq e^{-\lambda \tau }\frac{1}{3\pi }  \label{2.21}
\end{equation}
the absolute value of all coupling constants is bounded by one. The
branching process, being identical to a Galton-Watson process, terminates
with probability one and the number of inputs to the functional is finite
(with probability one). With the bounds on the coupling constants, the
multiplicative functional is bounded by one in absolute value almost surely.

Once a stochastic representation is obtained for $\chi \left( \xi _{1},\xi
_{2},\tau \right) $, one also has, by (\ref{2.8}), a stochastic
representation for the solution of the Fourier-transformed Poisson-Vlasov
equation and one obtains:

\textbf{Proposition 1.} \textit{The process I, above described, provides a
stochastic representation for the Fourier-transformed solutions of the
Poisson-Vlasov equation }$F_{i}\left( \xi _{1},\xi _{2},t\right) $\textit{\
for any arbitrary finite value of the arguments, provided the initial
conditions at time zero satisfy the boundedness conditions (A).}

So far we have constructed a general process that provides a stochastic
representation for the interacting Vlasov equation, not only for the Poisson
case, but also for more general situations with quadratic nonlinearities.
However, because of the integrated nature of the Coulomb interaction, the
Poisson case is special in that there is also a representation by a simpler
process. Looking at equation (\ref{2.9}) one sees that because of the factor 
$\stackrel{\rightarrow }{\xi _{2}}\cdot \stackrel{\wedge }{\xi
_{1}^{^{\prime }}}$ only the trees where the mode $\chi _{j}\left( \xi
_{1}^{^{\prime }},0,\tau -s\right) $ survives until time zero will
contribute to the functional. That is, the only trees with non-zero
contributions to the functional (\ref{2.17}) are the one-sided trees
represented in Fig.2. Therefore for the calculation of the solution one may
replace $\chi _{j}\left( \xi _{1}^{^{\prime }},0,\tau -s\right) $ by the
initial condition computed at $\left( \xi _{1}^{^{\prime }},\left( \tau
-s\right) \xi _{1}^{^{\prime }}\right) $. The process then becomes the
following linear process with random couplings 
\begin{eqnarray}
&&X_{i}^{\left( k+1\right) }\left( \xi _{1},\xi _{2},\tau \right)  \nonumber
\\
&=&\chi _{i}\left( \xi _{1},\xi _{2}+\tau \frac{\xi _{1}}{\gamma \left(
\left| \xi _{2}\right| \right) },0\right) \mathbf{1}_{\left[ s>\tau \right] }
\label{2.22} \\
&&+g_{ii}^{^{\prime }}\left( \xi _{1},\xi _{1}^{^{\prime }},s\right)
X_{i}^{\left( k\right) }\left( \xi _{1}-\xi _{1}^{^{\prime }},\xi _{2}+s%
\frac{\xi _{1}}{\gamma \left( \left| \xi _{2}\right| \right) },\tau
-s\right) \mathbf{1}_{\left[ s<\tau \right] }\mathbf{1}_{\left[
l_{s}=0\right] }  \nonumber \\
&&+g_{ij}^{^{\prime }}\left( \xi _{1},\xi _{1}^{^{\prime }},s\right)
X_{i}^{\left( k\right) }\left( \xi _{1}-\xi _{1}^{^{\prime }},\xi _{2}+s%
\frac{\xi _{1}}{\gamma \left( \left| \xi _{2}\right| \right) },\tau
-s\right) \mathbf{1}_{\left[ s<\tau \right] }\mathbf{1}_{\left[
l_{s}=1\right] }  \nonumber
\end{eqnarray}
the coupling constant at the branchings being 
\begin{equation}
g_{ij}^{^{\prime }}\left( \xi _{1},\xi _{1}^{^{\prime }},s\right) =-\frac{%
8\pi e_{i}e_{j}}{m_{i}\lambda }\frac{\left( \left| \xi _{1}^{^{\prime
}}\right| ^{-1}h*h\right) \left( \xi _{1}\right) }{h\left( \xi _{1}\right) }%
\frac{\stackrel{\rightarrow }{\xi _{2}}\cdot \stackrel{\wedge }{\xi
_{1}^{^{\prime }}}}{\gamma \left( \left| \xi _{2}\right| \right) }\chi
_{j}\left( \xi _{1}^{^{\prime }},\left( \tau -s\right) \xi _{1}^{^{\prime
}},0\right)  \label{2.23}
\end{equation}

\begin{figure}[tbh]
\begin{center}
\psfig{figure=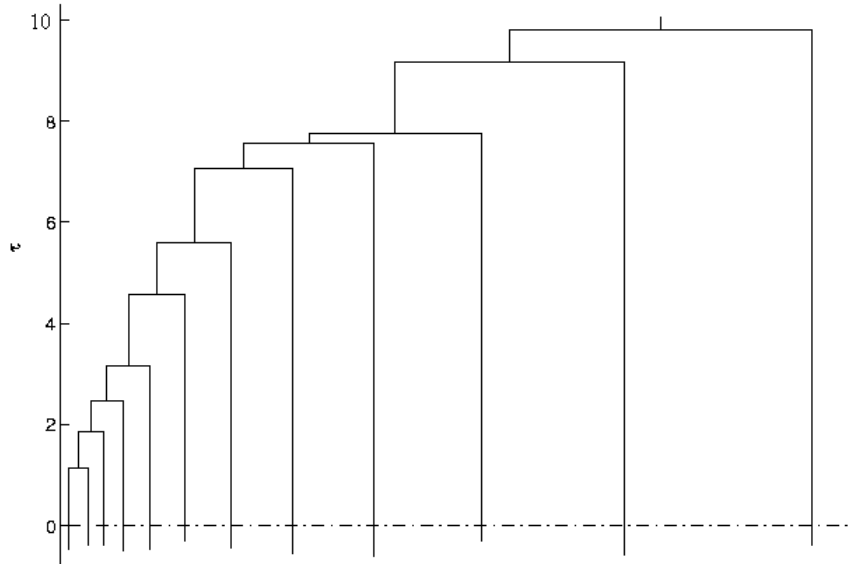,width=11truecm}
\end{center}
\caption{A one-sided tree corresponding to process II}
\end{figure}

The functional representing the solution is the product of all branching
coupling constants times one additional factor corresponding to the last
non-branching mode. The result is the following

\textbf{Proposition 2.} \textit{The linear process II, defined by (\ref{2.22}%
) and (\ref{2.23}) also provides a stochastic representation of the
solutions of the Poisson-Vlasov equation, the conditions on the kernels and
initial conditions being given by (A), (\ref{2.12}) and (\ref{2.20}).}

\section{Stochastic representation and numerical codes}

The backwards-in-time probabilistic representations, obtained in Sect.2,
seem appropriate for the numerical evaluation of Fourier-localized
solutions. Good statistics requires the average of the multiplicative
functional over many realization trees. In the backwards in time realization
one fixes a particular mode at time $\tau $ and generates as many trees as
needed for that particular mode. Notice that by studying high Fourier modes
one may obtain information about the small scale behaviour of the solution
without having the need for a fine grid as it would be necessary in a real
space numerical code. Each realization tree being independent of all the
others, the probabilistic code is also appropriate for parallelization.

We will not report, in this paper, extensive calculations using these
representations and the corresponding codes. Nevertheless we list all the
needed probability distributions needed to implement the method.

For the construction of the sample trees and the calculation of the
functional, the following probability densities are needed:

- The probability of a $\xi _{1}-$ mode branching into $\xi _{1}^{^{\prime
}} $ and $\xi _{1}-\xi _{1}^{^{\prime }}$ modes 
\begin{equation}
p\left( \xi _{1},\xi _{1}^{^{\prime }}\right) =\frac{\left| \xi
_{1}^{^{\prime }}\right| ^{-1}\left( 1+\left| \xi _{1}-\xi _{1}^{^{\prime
}}\right| ^{2}\right) ^{-2}\left( 1+\left| \xi _{1}^{^{\prime }}\right|
^{2}\right) ^{-2}}{\Gamma \left( \left| \xi _{1}\right| \right) }
\label{3.1}
\end{equation}
with $\Gamma \left( \xi _{1}\right) $ given by Eq.(\ref{2.13}). One notices
that, for each $\left| \xi _{1}\right| $, this probability is only function
of two variables, the $\left| \xi _{1}^{^{\prime }}\right| $ and the angle
between $\xi _{1}$ and $\xi _{1}^{^{\prime }}$. Therefore defining 
\begin{equation}
z=\frac{1}{1+\left| \xi _{1}^{^{\prime }}\right| ^{2}}  \label{3.2}
\end{equation}
and changing the integration measure one obtains a

- Probability density $p\left( z,\cos \theta \right) $%
\begin{equation}
p\left( z,\cos \theta \right) =\frac{\pi }{\Gamma \left( \left| \xi
_{1}\right| \right) }\frac{1}{\left( \left| \xi _{1}\right| ^{2}+\frac{1}{z}%
-2\left| \xi _{1}\right| \cos \theta \sqrt{\frac{1}{z}-1}\right) ^{2}}
\label{3.3}
\end{equation}
with $z$ in the interval $\left( 0,1\right) $ and $\cos \theta $ in the
interval $\left( -1,1\right) $.

Because the inverse of the cumulative distribution functions have not a nice
analytic form, we may use the reject method in the plane $\left( z,\cos
\theta \right) $ to simulate this probability distribution. For this, one
needs 
\begin{equation}
p\left( z,\cos \theta \right) _{\max }=\frac{\pi }{\Gamma \left( \left| \xi
_{1}\right| \right) }  \label{3.4}
\end{equation}
which is obtained for $\cos ^{2}\theta =1$ and $\frac{1}{z}=\left| \xi
_{1}\right| ^{2}+1$.

However because $p\left( z,\cos \theta \right) _{\max }$ is very large for
large $\left| \xi _{1}\right| $ in a narrow region, it is more efficient to
use

- The integrated $p\left( z\right) $ density, 
\begin{equation}
p\left( z\right) =\frac{2\pi }{\Gamma \left( \left| \xi _{1}\right| \right) }%
\frac{1}{\left( \left| \xi _{1}\right| ^{2}-\frac{1}{z}\right) ^{2}+4\left|
\xi _{1}\right| ^{2}}  \label{3.5}
\end{equation}
with $p\left( z\right) _{\max }=p\left( \min \left( \left| \xi _{1}\right|
^{-2},1\right) \right) $, to choose $z$ and then, once $z$ is chosen, to use

- The conditional probability density $p\left( \cos \theta |z\right) $%
\begin{equation}
p\left( \cos \theta |z\right) =\frac{\left( \left| \xi _{1}\right| ^{2}-%
\frac{1}{z}\right) ^{2}+4\left| \xi _{1}\right| ^{2}}{2\left( \left| \xi
_{1}\right| ^{2}+\frac{1}{z}-2\left| \xi _{1}\right| \cos \theta \sqrt{\frac{%
1}{z}-1}\right) ^{2}}  \label{3.6}
\end{equation}
with 
\begin{equation}
\max_{\theta }p\left( \cos \theta |z\right) =\frac{\left( \left| \xi
_{1}\right| ^{2}-\frac{1}{z}\right) ^{2}+4\left| \xi _{1}\right| ^{2}}{%
2\left( \left| \xi _{1}\right| ^{2}+\frac{1}{z}-2\left| \xi _{1}\right| 
\sqrt{\frac{1}{z}-1}\right) ^{2}}  \label{3.7}
\end{equation}
to choose $\cos \theta $.

Once $z$ and $\cos \theta $ are chosen, one computes $\left| \xi
_{1}^{^{\prime }}\right| =\sqrt{\frac{1}{z}-1}$ and chooses 
\[
\varphi =2\pi RAND 
\]
RAND being a random variable uniformly distributed in the interval $\left(
0,1\right) $. Then one obtains 
\[
\xi _{1}^{^{\prime }}=\left| \xi _{1}^{^{\prime }}\right| \left( \sin \theta
\cos \varphi ,\sin \theta \sin \varphi ,\cos \theta \right) 
\]
and $\xi _{1}-\xi _{1}^{^{\prime }}$.

Notice that only the amplitudes of the Fourier modes $\left| \xi
_{1}^{^{\prime }}\right| $ and $\left| \xi _{1}-\xi _{1}^{^{\prime }}\right| 
$ are needed as inputs to compute the probabilities in the next branchings
but the full vector is needed to compute $\xi _{2}$. Finally the lifetime $%
\tau $ of each mode is obtained from 
\begin{equation}
\tau =\frac{\ln \left( RAND\right) }{\lambda }  \label{3.9}
\end{equation}

For the trees a standard indexation is used, each tree being a row vector of
integer numbers with the number $k$ at the position $n$ meaning that that
mode was born at the branching of the mode in the position $k$.

The conditions (A), (\ref{2.12}) and (\ref{2.20}) guarantee that all factors
entering the multiplicative functional (\ref{2.17}) are bounded by one,
implying that the functional itself is also bounded. This, together with the
Galton-Watson nature of the branching, insures convergence of the
expectation value. However, in practice, this leads to very small values of
the functional and for large times (large trees) one may be faced with
round-off inaccuracies in the computer. In fact the limitation to factors
strictly not larger than one is only imposed for mathematical convenience.
What is actually needed for convergence is that the functional be bounded by
some value with probability one. A more relaxed condition on the constants
may therefore be obtained by imposing $\left| p_{n}g_{\max }^{n}\left(
I\right) F_{\max }^{n+1}\right| <M$ for process I and $\left| p_{n}g_{\max
}^{n}\left( II\right) F_{\max }\right| <M$ for process II, $p_{n}$ being the
probability of a tree with $N$ branchings, $g_{\max }^{n}\left( I\right) $
and $g_{\max }^{n}\left( II\right) $ the maximum values of the couplings and 
$F_{\max }$ the maximum value of the initial condition.

To test the method we have studied the time evolution of small and large
Fourier modes in a plasma with two particle species of opposite charges, one
light and the other heavy, with two types of initial Fourier distribution
functions, namely 
\begin{equation}
F_{i}^{(1)}\left( \xi _{1},\xi _{2},0\right) =C_{0}^{(1)}e^{-\gamma \left|
\xi _{1}\right| ^{2}}e^{-\beta _{i}\left| \xi _{2}\right| ^{2}}  \label{3.10}
\end{equation}
and 
\begin{equation}
\begin{array}{lll}
F_{+}^{(2)}\left( \xi _{1},\xi _{2},0\right) & = & C_{0+}^{(2)}e^{-\gamma
\left| \xi _{1}\right| ^{2}}e^{-\beta _{+}\left| \xi _{2}\right| ^{2}} \\ 
F_{-}^{(2)}\left( \xi _{1},\xi _{2},0\right) & = & C_{0-}^{(2)}e^{-\gamma
\left| \xi _{1}\right| ^{2}}\theta \left( k-\left| \xi _{2}\right|
^{2}\right)
\end{array}
\label{3.10a}
\end{equation}
$\beta _{+}=40\beta _{-}$ and the $C_{0}^{(i)\prime }s$ are chosen to
fulfill condition (A). For the initial condition we have chosen $\xi
_{1}=\left( 0.01,0.01,0.01\right) $ and varied $\left| \xi _{2}\right| ^{2}$
in the range $3.10^{-4}$ to $7$. Then, the time evolution is computed using
the one-sided representation (process II). Some results are shown in Figs.3
and 4, with time in units of $\frac{1}{\lambda }$. Although it is known that
on $\mathbb{R}^{3}$ and without an external force there are no nontrivial
steady states when both charges of opposite sign can move\cite{Rein1} \cite
{Rein2}, one can see the relative stability of the Fourier modes for short
times for the Gaussian initial condition $F^{\left( 1\right) }$, whereas for
the $F^{\left( 2\right) }$ initial condition one sees the appearance of
growing Fourier modes that were not present in the initial density. Although
the points were computed for the same set of final $\tau ^{\prime }s$, in
the plots we show the actual time $t$, obtained from (\ref{2.5}).

\begin{figure}[tbh]
\begin{center}
\psfig{figure=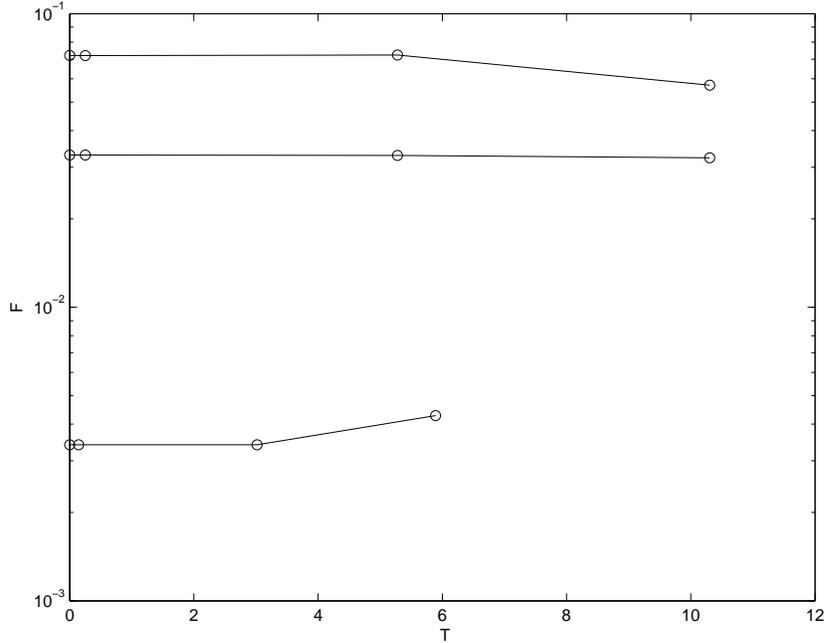,width=11truecm}
\end{center}
\caption{Time evolution of some Fourier modes for the $F^{\left( 1\right)}$
initial condition, $T=\lambda t$}
\end{figure}

\begin{figure}[tbh]
\begin{center}
\psfig{figure=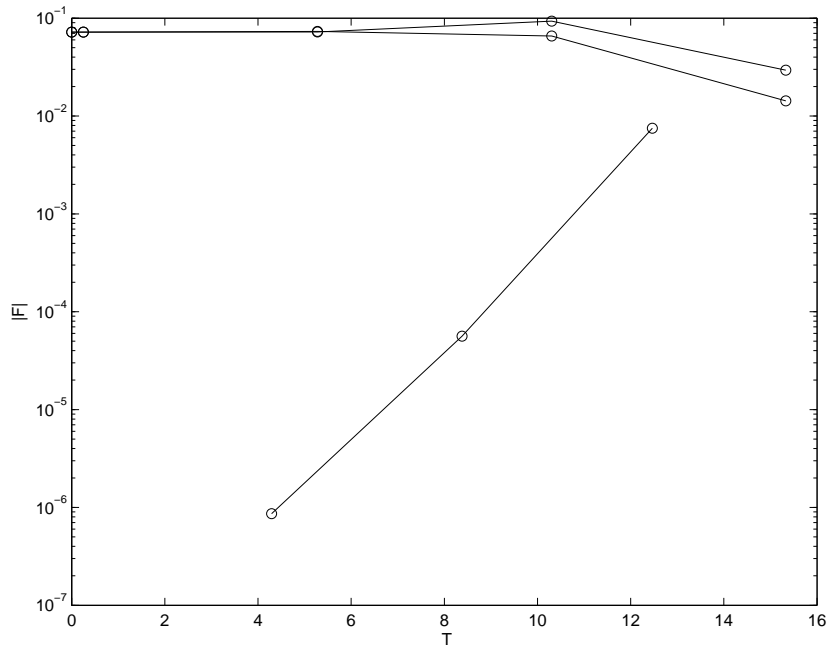,width=11truecm}
\end{center}
\caption{Time evolution of some Fourier modes for the $F^{\left( 2\right)}$
initial condition, $T=\lambda t$}
\end{figure}

Notice that to obtain a reasonable stability of the averaged functional one
needs to compute many sample trees. The points shown in the figures were
obtained with averages over $10^{6}$ or $2\times 10^{6}$ trees, depending on
the evolution time. The reason for the need for a large number of sample
trees arises from the fact that, for large times, most trees contribute a
very small value to the average, the actual average arising from the
contribution of a small number of them. This calls for the need to control
the results by a large deviation analysis (see below).

\section{Conclusions}

1 - Mostly when localized solutions in Fourier space (or in configuration
space for other stochastic representations) are desired, the method seems
appropriate. When a global calculation of the solution is desired, the
stochastic representation method is probably not competitive with other
current simulation methods. The computational example presented in Sect. 3,
merely illustrative of the method, was obtained with modest computational
means. Our purpose was mostly to test the stability of the results. To
obtain good statistics and also to study the fluctuation spectrum of the
process, many sample trees have to be used for each initial condition.
However, because each tree is independent from the others and also because
after the branching each mode evolves independently of the others, this
algorithm is well suited for parallelization and distributed computing. In
this sense the stochastically-based algorithms might also become competitive
even for global calculations using parallel computing. In fact stochastic
representations have already been found to be efficient for domain
decomposition in parallel computing \cite{Acebron1} \cite{Acebron2}.

2 - The fluctuations around the mean in a branching process are typically
very much non-Gaussian. Therefore a simple calculation of the standard
deviation or other lower order momenta are not sufficient to check the
reliability of the results. A large deviation analysis is recommended for
numerical calculations using branching processes. Some general results on
large deviations in branching processes are known\cite{Biggins} \cite
{Athreya} \cite{Ney} \cite{Wachtel}. Of more practical importance are
probably methods to estimate large deviation effects directly from the data.
This may be done, for example, by the empirical construction of the
deviation function. This is done by the empirical construction of the free
energy and from it, by Legendre transform, the deviation function. For
details we refer to \cite{Vilela2}. Given a deviation function $I\left(
x\right) $, the probability of obtaining a value $x$ for the empirical
average of a sample of size $n$ is 
\[
P_{n}\left( dx\right) \asymp e^{-nI\left( x\right) }dx 
\]
where $\asymp $ means logarithmic equivalence. We have used the method
described in \cite{Vilela2} to check the reliability of the results. In the
Fig.5 we present the empirically obtained deviation function for a sample of 
$5\times 10^{5}$ trees. At first sight the regular behavior of $I\left(
y\right) $ around the mean, seen in the upper plot of Fig.5, would seem to
indicate that the distribution is Gaussian. However expanding a little more
(in the lower plot) the domain of the variable $x$ one sees the very
non-Gaussian nature of the data. It means that, had we used a smaller
sample, any empirical mean in the range $0.04-0.075$ would have been likely.
A rough lower bound on the size of the needed sample may be obtained from
the inverse of the deviation function at the point where the behavior of $%
I\left( x\right) $ changes.

\begin{figure}[tbh]
\begin{center}
\psfig{figure=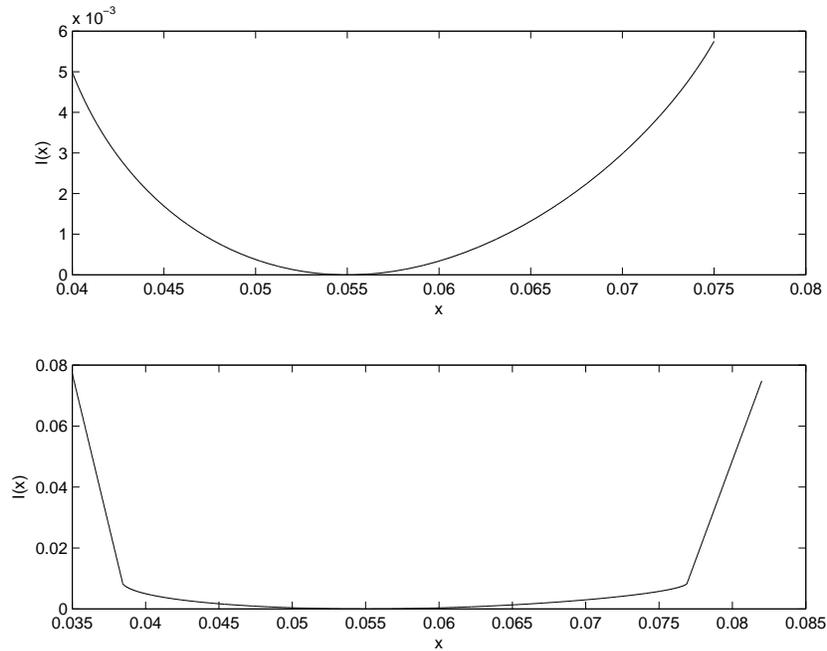,width=11truecm}
\end{center}
\caption{The behavior of the deviation function for a sample of size $%
5\times 10^{5}$}
\end{figure}

3 - Stochastic representations of the solutions of deterministic equations
may have some relevance for the study of the fluctuation spectrum. In the
past, the fluctuation spectrum of charged fluids was studied either by the
BBGKY hierarchy derived from the Liouville or Klimontovich equations, with
some sort of closure approximation, or by direct approximations to the
N-body partition function or by models of dressed test particles, etc. (see
reviews in \cite{Oberman} \cite{Krommes}). Alternatively, by linearizing the
Vlasov equation about a stable solution and diagonalizing the Hamiltonian, a
canonical partition function may be used to compute correlation functions 
\cite{Morrison}.

As a model for charged fluids, the Vlasov equation is just a mean-field
collisionless theory. Therefore, it is unlikely that, by itself, it will
contain full information on the fluctuation spectrum. Kinetic and fluid
equations are obtained from the full particle dynamics in the 6N-dimensional
phase-space by a chain of reductions. Along the way, information on the
actual nature of fluctuations and turbulence may have been lost. An accurate
model of turbulence may exist at some intermediate (mesoscopic) level, but
not necessarily in the final mean-field equation.

When a stochastic representation is constructed, one obtains a process for
which the mean value is the solution of the mean-field equation. The process
itself contains more information. This does not mean, of course, that the
process is an accurate mesoscopic model of Nature, because we might be
climbing up a path different from the one that led us down from the particle
dynamics. Nevertheless, insofar as the stochastic representation is
qualitatively unique and related to some reasonable iterative process, it
provides a surrogate mesoscopic model from which fluctuations are easily
computed. This is what we have referred elsewhere as \textit{the stochastic
principle }\cite{Vilela1}. At the minimum, one might say that the stochastic
principle provides yet another closure procedure for the kinetic equations.

\end{document}